\documentclass[12pt]{article}
\usepackage{epsf}
\usepackage[dvips]{color}
\usepackage{exscale}
\usepackage[dvips]{graphicx}
\usepackage{amsmath}
\usepackage{amssymb}
\usepackage{latexsym}

\begin{document}
\setlength{\baselineskip}{18pt}
\begin{titlepage}
\begin{flushright}
\begin{tabular}{l}
EPHOU-13-006
\end{tabular} 
\end{flushright}

\vspace*{1.2cm}
\begin{center}
{\Large\bf Renormalization group invariants in neutrino sector}
\end{center}
\lineskip .75em
\vskip 1.5cm

\begin{center}
{\large Naoyuki Haba$^{1,2}$ and Ryo Takahashi$^2$}\\

\vspace{1cm}

$^1${\it Graduate School of Science and Engineering, Shimane University, 

Matsue 690-8504, Japan}\\

$^2${\it Department of Physics, Faculty of Science, Hokkaido University, 

Sapporo 060-0810, Japan}\\

\vspace*{10mm}
{\bf Abstract}\\[5mm]
{\parbox{13cm}{\hspace{5mm}
We show renormalization group invariants in neutrino sector. These are found from a 
simple analytical discussion of Majorana mass matrix for light neutrinos. It is 
shown that the invariants are obtained by taking ratios among elements of the 
Majorana neutrino mass matrix. The invariance is independent of neutrino mass 
ordering and a parameterization of mixing matrix for the lepton sector. 
Parameters in the runnings under renormalization group equations in the neutrino
 sector are also analyzed.}}
\end{center}
\end{titlepage}

\section{Introduction}
Neutrino physics is one of important clues to look for physics beyond the 
standard model (SM) because neutrino oscillation experiments established that 
neutrinos have small masses compared to other SM fermion masses. Moreover recent
 precision measurements of leptonic mixing angles in the 
Pontecorvo-Maki-Nakagawa-Sakata (PMNS) matrix~\cite{Maki:1962mu} showed that 
$\theta_{12}$ and $\theta_{23}$ are large, and $\theta_{13}$ is small but not 
zero~\cite{An:2012eh,tortola,GonzalezGarcia:2012sz}. As far as the neutrino 
masses concerned, two mass squared differences are measured. Accordingly, a 
normal hierarchy (NH), $m_1<m_2<m_3$, or an inverted hierarchy (IH), 
$m_3<m_1<m_2$, are now allowed where $m_i$ are the neutrino mass eigenvalues. 

It is, in particular, important to realize the smallness of the neutrino masses.
 A lot of works have been proposed for the realization of the small masses in 
the context of the seesaw mechanism~\cite{seesaw}. The mechanism implies the 
presence of new physics (particles) at a high energy regime. In addition to the 
seesaw mechanism, one may also consider a kind of unified theory, namely a grand
 unified theory (GUT), and/or a new symmetry, e.g., flavor symmetry, at high 
energy in order to nicely derive a gauge (group) and Yukawa (masses and 
generation mixing) structures of the SM at low energy. Therefore, investigations
 of high energy behavior in the neutrino sector might be important for a 
clarification of a new physics beyond the SM. Renormalization of group equations
 (RGEs) for physical parameters are ones of tools to seek a high energy physics.
 In fact, a large number of works respect with the RGEs of the neutrino sector 
have been presented (e.g., 
see~\cite{Chankowski:1993tx,Ellis:1999my,Haba:1998fb,Haba:1999ca,Haba:1999fk,Haba:2012ar}). 

Supersymmetry (SUSY) is attractive since it can solve some problems in the SM. 
For instance, the problems are the gauge hierarchy problem and the absence of 
the dark matter (DM) candidate. Furthermore, the gauge coupling unification can 
be interestingly realized at high energy scale in the minimal supersymmetric 
standard model (MSSM). Then, the origin of the tiny neutrino masses, e.g. the 
heavy right-handed neutrinos, could be naturally embedded into a GUT. In this 
letter, we investigate a behavior of a coefficient of the Weinberg 
operator~\cite{Weinberg:1979sa}, which describes tiny neutrino mass, under the 
RGEs in the MSSM.

\section{Analyses under renormalization group equations in neutrino sector}

\subsection{Renormalization group invariants}

We consider effective Yukawa interactions of the lepton sector and the Weinberg
 operator in the MSSM at low energy such as the electroweak (EW) scale 
$\Lambda_{\rm EW}$ in the lepton sector,
 \begin{eqnarray}
  \mathcal{L}\supset-y_e\bar{L}H_de_R+\frac{\kappa}{2}(H_uL)(H_uL)+h.c.,
 \end{eqnarray}
where $y_e$, $L$, $e_R$, $H_{d(u)}$, and the second term are a matrix of Yukawa
 couplings of charged leptons, left-handed lepton doublets, right-handed 
charged leptons, down(up)-type Higgs, and the Weinberg operator, respectively. 
A mass matrix for light (active) neutrinos is given by 
$M_\nu=\kappa v_u^2=\kappa v^2\sin^2\beta$ after the Higgs gets a vacuum 
expectation value (VEV), $v_u$ where $\tan\beta\equiv v_u/v_d$ and $v_d$ is a 
VEV of the down-type Higgs.
 
 The light neutrino mass matrix can also be described by the PMNS matrix and mass eigenvalues of light neutrinos as:
  \begin{eqnarray}
   (M_\nu)_{\alpha\beta}=(UM_\nu^{\rm diag}U^T)_{\alpha\beta}=(U\cdot\mbox{Diag}\{m_1,m_2,m_3\}\cdot U^T)_{\alpha\beta}=\sum_iU_{\alpha i}U_{\beta i}m_i, \label{lM}
  \end{eqnarray}
in a diagonal basis of the Yukawa coupling matrix for the charged leptons where 
$U$ is the PMNS matrix, $M_\nu^{\rm diag}$ is a diagonal matrix, and 
$\alpha,\beta=e,\mu,\tau$. The light neutrino mass matrix can be described by 
3 mixing angles, 3 mass eigenvalues of the neutrinos, and 3 CP-phases (one Dirac
 and two Majorana phases) if the neutrinos are Majorana particles. To determine 
these quantities in the neutrino sector by the experiments is one of important 
goals in studies of neutrino physics. Once one fixes those values at low energy 
as boundary conditions, one can obtain those values at arbitrary high energy 
scale by solving corresponding RGEs. To get a structure of neutrino mass matrix 
with ones of other SM fermion mass matrices strongly motivates to study physics 
beyond the SM, e.g., GUT and/or the presence of an additional symmetry such as a
 flavor symmetry.

By solving the RGE for $\kappa$ as:
 \begin{eqnarray}
  16\pi^2\frac{d\kappa}{dt}=\bar{\alpha}\kappa+(y_ey_e^\dagger)\kappa
                            +\kappa(y_ey_e^\dagger)^T,
 \end{eqnarray}
with $t\equiv\ln\mu$ and
 \begin{eqnarray}
  \bar{\alpha}\equiv-\frac{6}{5}g_1^2-6g_2^2+6(y_u^2+y_c^2+y_t^2), \label{al}
 \end{eqnarray}
at one-loop level, we can write a Majorana mass matrix of the light neutrinos 
as $M_\nu(\Lambda)=R(IM_\nu(\Lambda_{\rm EW})I)$ at arbitrary high energy scale 
$\Lambda$ where $\mu$ is a renormalization scale, $g_i$ are gauge coupling 
constants, $y_{\alpha_q}$ ($\alpha_q=u,c,t$) are Yukawa couplings of up-type 
quarks, $R$ is a flavor blind overall factor, and $I$ is defined by 
$I^{-1}\equiv\mbox{Diag}\{\sqrt{I_e},\sqrt{I_\mu},\sqrt{I_\tau}\}$~\cite{Ellis:1999my,Haba:1998fb,Haba:1999fk}. 
$I_\alpha$ denote quantum corrections for the Yukawa couplings as 
$I_\alpha\equiv\mbox{exp}\left[\frac{1}{8\pi^2}\int_{t_\Lambda}^{t_{\rm 
EW}}dty_\alpha^2\right]$ with $t_\Lambda\equiv\ln\Lambda$ and $t_{\rm 
EW}\equiv\ln\Lambda_{\rm EW}$. A dominant effect of the quantum corrections 
comes from the coupling $y_\tau$, and thus we introduce small parameters 
defined as $\epsilon_e\equiv\sqrt{\frac{I_\tau}{I_e}}-1$ and 
$\epsilon_\mu\equiv\sqrt{\frac{I_\tau}{I_\mu}}-1$. Since $\epsilon_e$ and 
$\epsilon_\mu$ are numerically well approximated as $\epsilon_e=\epsilon_\mu$, 
we take $\epsilon=\epsilon_e=\epsilon_\mu$ in the following discussions. Then, 
the neutrino mass matrix at arbitrary high energy scale can be well approximated
 by
 \begin{eqnarray}
  M_\nu(\Lambda)=r 
   \left(
    \begin{array}{ccc}
     (M_\nu(\Lambda_{\rm EW}))_{ee} & (M_\nu(\Lambda_{\rm EW}))_{e\mu} & (M_\nu(\Lambda_{\rm EW}))_{e\tau}(1+\epsilon) \\
     (M_\nu(\Lambda_{\rm EW}))_{e\mu} & (M_\nu(\Lambda_{\rm EW}))_{\mu\mu} & (M_\nu(\Lambda_{\rm EW}))_{\mu\tau}(1+\epsilon) \\
     (M_\nu(\Lambda_{\rm EW}))_{e\tau}(1+\epsilon) & (M_\nu(\Lambda_{\rm EW}))_{\mu\tau}(1+\epsilon) & (M_\nu(\Lambda_{\rm EW}))_{\tau\tau}(1+\epsilon)^2
    \end{array}
   \right), \label{hM}
 \end{eqnarray}
where $r\equiv R/I_e$.

Let us consider renormalization group invariants for the neutrino parameters. 
First, it has been pointed out that argument of all the matrix elements of 
$\kappa$ do not evolve under the RGEs~\cite{Haba:1999ca}. The reason is as 
follows: the RGE of $\kappa$ can be rewritten as,
 \begin{eqnarray}
  \frac{d}{dt}\ln\kappa_{ij}
  =\frac{d}{dt}\ln\left|\kappa_{ij}\right|+i\frac{d}{dt}\phi_{ij}
  =\gamma_i+\gamma_j+\gamma_H,
 \end{eqnarray}
with a notation $\kappa_{ij}\equiv|\kappa_{ij}|e^{i\phi_{ij}}$ where $\gamma_i$ 
and $\gamma_H$ are anomalous dimensions, which are real, for the left-handed 
lepton doublets and the up-type Higgs defined by those wave function 
renormalizations, respectively. Therefore, this leads to $d\phi_{ij}/dt=0$. Are 
there any other renormalization group invariants in addition to CP-phases in 
$\kappa$?

When one compare \eqref{hM} with \eqref{lM}, one can interestingly find 4 
renormalization group invariants as,\footnote{See also \cite{Mohapatra:2006xy} 
for a property of a strong scaling ansatz under the RGEs.}
 \begin{eqnarray}
  && \frac{(M_\nu(\Lambda))_{ee}}{(M_\nu(\Lambda))_{e\mu}} 
     =\frac{(M_\nu(\Lambda_{\rm EW}))_{ee}}{(M_\nu(\Lambda_{\rm EW}))_{e\mu}} 
     =\frac{\sum_iU_{ei}^2(\Lambda_{\rm EW})m_i(\Lambda_{\rm EW})}
           {\sum_iU_{ei}(\Lambda_{\rm EW})U_{\mu i}(\Lambda_{\rm EW})
            m_i(\Lambda_{\rm EW})}
     =\frac{m_{ee}}{m_{e\mu}}, \\
  && \frac{(M_\nu(\Lambda))_{ee}}{(M_\nu(\Lambda))_{\mu\mu}} 
     =\frac{(M_\nu(\Lambda_{\rm EW}))_{ee}}{(M_\nu(\Lambda_{\rm EW}))_{\mu\mu}} 
     =\frac{\sum_iU_{ei}^2(\Lambda_{\rm EW})m_i(\Lambda_{\rm EW})}
           {\sum_iU_{\mu i}^2(\Lambda_{\rm EW})m_i(\Lambda_{\rm EW})}
     =\frac{m_{ee}}{m_{\mu\mu}}, \\
  && \frac{(M_\nu(\Lambda))_{e\tau}}{(M_\nu(\Lambda))_{\mu\tau}} 
     =\frac{(M_\nu(\Lambda_{\rm EW}))_{e\tau}}
           {(M_\nu(\Lambda_{\rm EW}))_{\mu\tau}} 
     =\frac{\sum_iU_{ei}(\Lambda_{\rm EW})U_{\tau i}(\Lambda_{\rm EW})
            m_i(\Lambda_{\rm EW})}
           {\sum_iU_{\mu i}(\Lambda_{\rm EW})U_{\tau i}(\Lambda_{\rm EW})
            m_i(\Lambda_{\rm EW})}
     =\frac{m_{e\tau}}{m_{\mu\tau}}, \\
  && \frac{(M_\nu(\Lambda))_{e\tau}^2}
          {(M_\nu(\Lambda))_{ee}(M_\nu(\Lambda))_{\tau\tau}} 
     =\frac{(M_\nu(\Lambda_{\rm EW}))_{e\tau}^2}
           {(M_\nu(\Lambda_{\rm EW}))_{ee}(M_\nu(\Lambda_{\rm EW}))_{\tau\tau}} 
     \nonumber \\
  && =\frac{\left(\sum_iU_{ei}(\Lambda_{\rm EW})U_{\tau i}(\Lambda_{\rm EW})
            m_i(\Lambda_{\rm EW})\right)^2}
           {\left(\sum_iU_{ei}^2(\Lambda_{\rm EW})m_i(\Lambda_{\rm EW})\right)
            \left(\sum_iU_{\tau i}^2(\Lambda_{\rm EW})m_i(\Lambda_{\rm EW})
            \right)}
     =\frac{m_{e\tau}^2}{m_{ee}m_{\tau\tau}}. 
 \end{eqnarray}
We show values of these invariants in Figures~\ref{fig1} and~\ref{fig1-2}. Figures~\ref{fig1} and~\ref{fig1-2} are cases that all CP-phases are relatively small ($0\leq(\delta,\alpha,\beta)<2\pi/3$) and large ($4\pi/3\leq(\delta,\alpha,\beta)<2\pi$), respectively, where $\delta$, $\alpha$ and $\beta$ are a Dirac and two Majorana phases, respectively. The calculations are given within $3\sigma$ range for experimentally determined values of the neutrino parameters~\cite{GonzalezGarcia:2012sz} as: 
 \begin{figure}
 \begin{center}
 \includegraphics[scale=0.3]{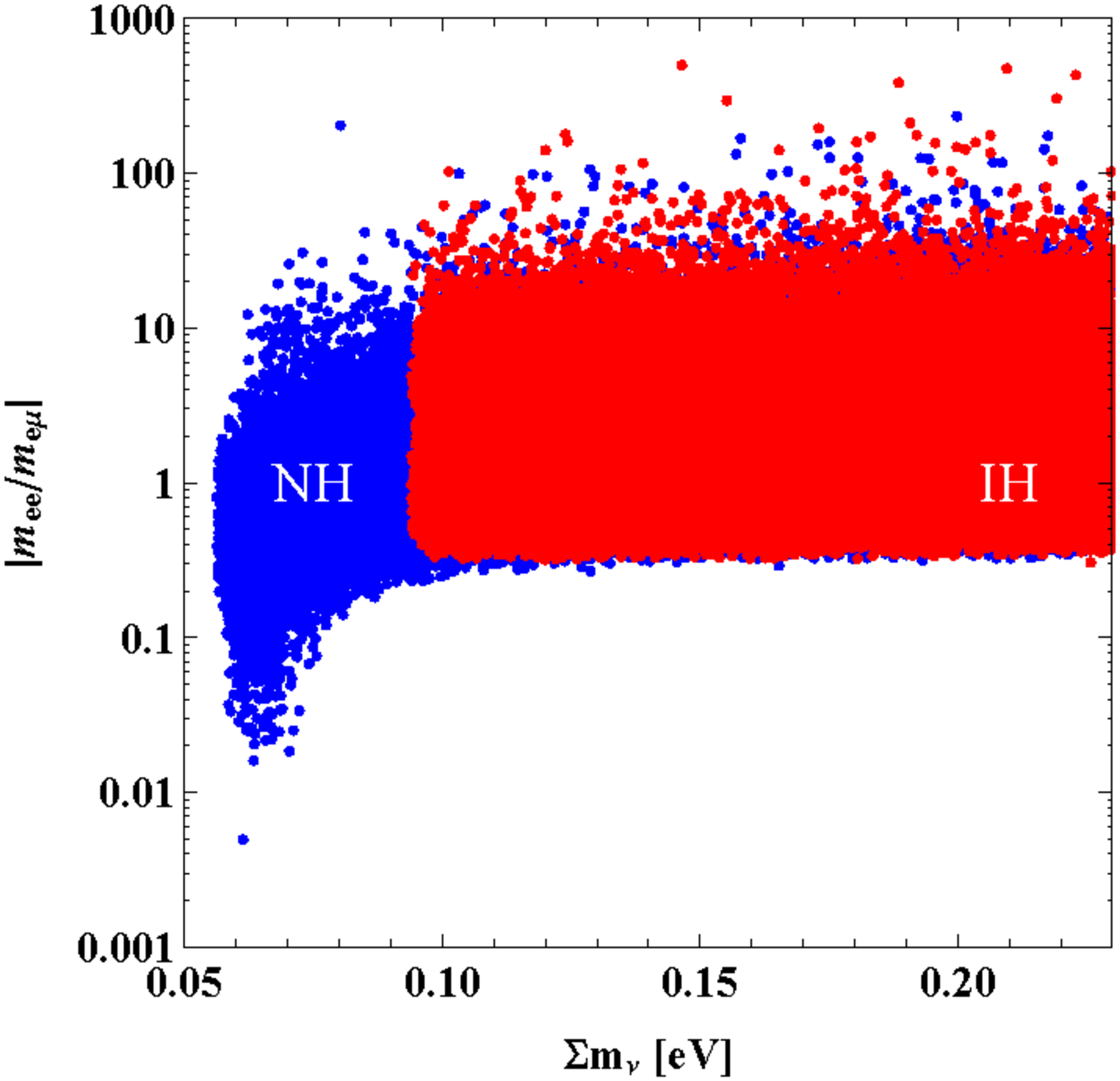}\hspace{8mm}
 \includegraphics[scale=0.3]{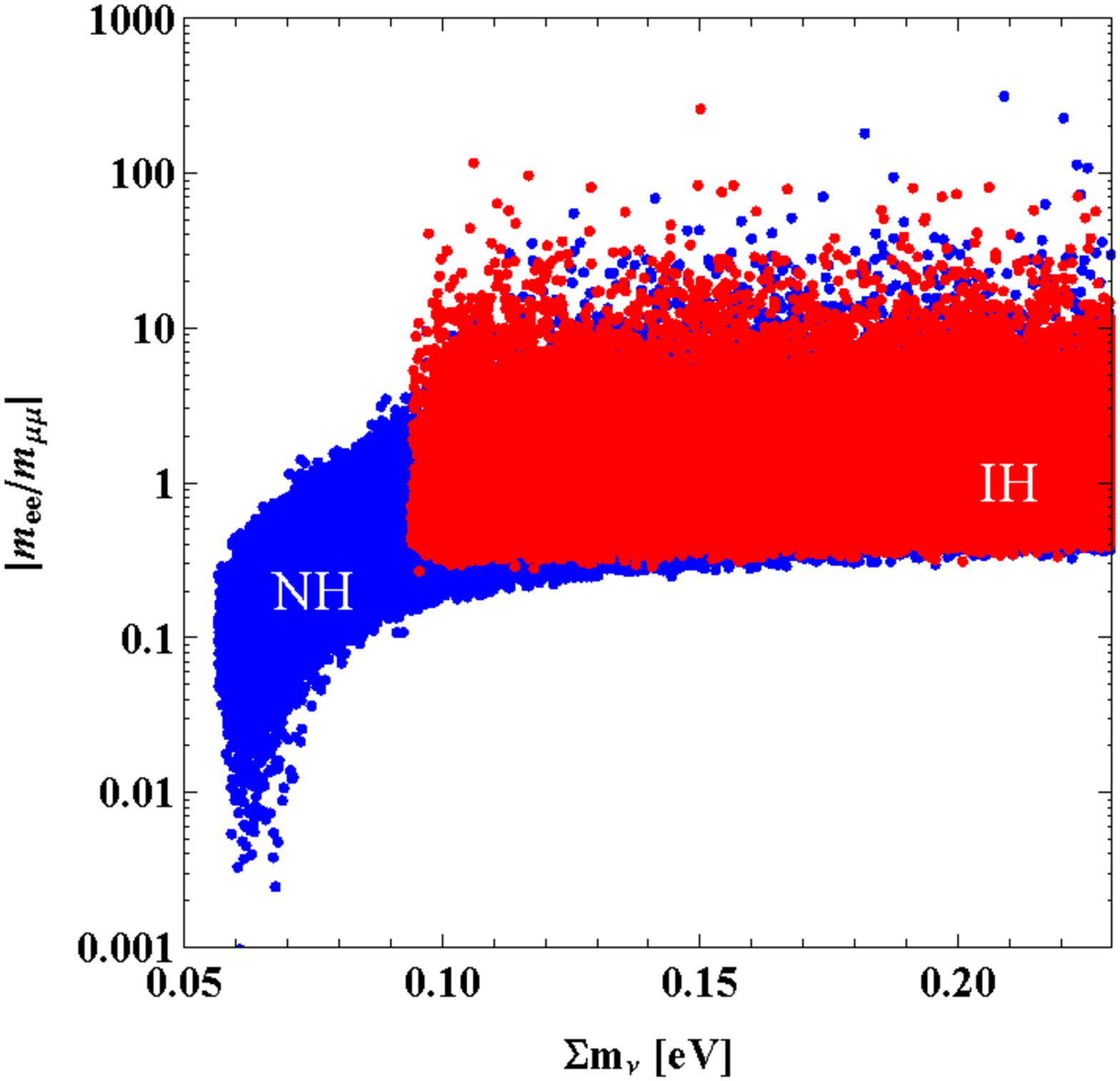}

%
~\\

%
 \includegraphics[scale=0.3]{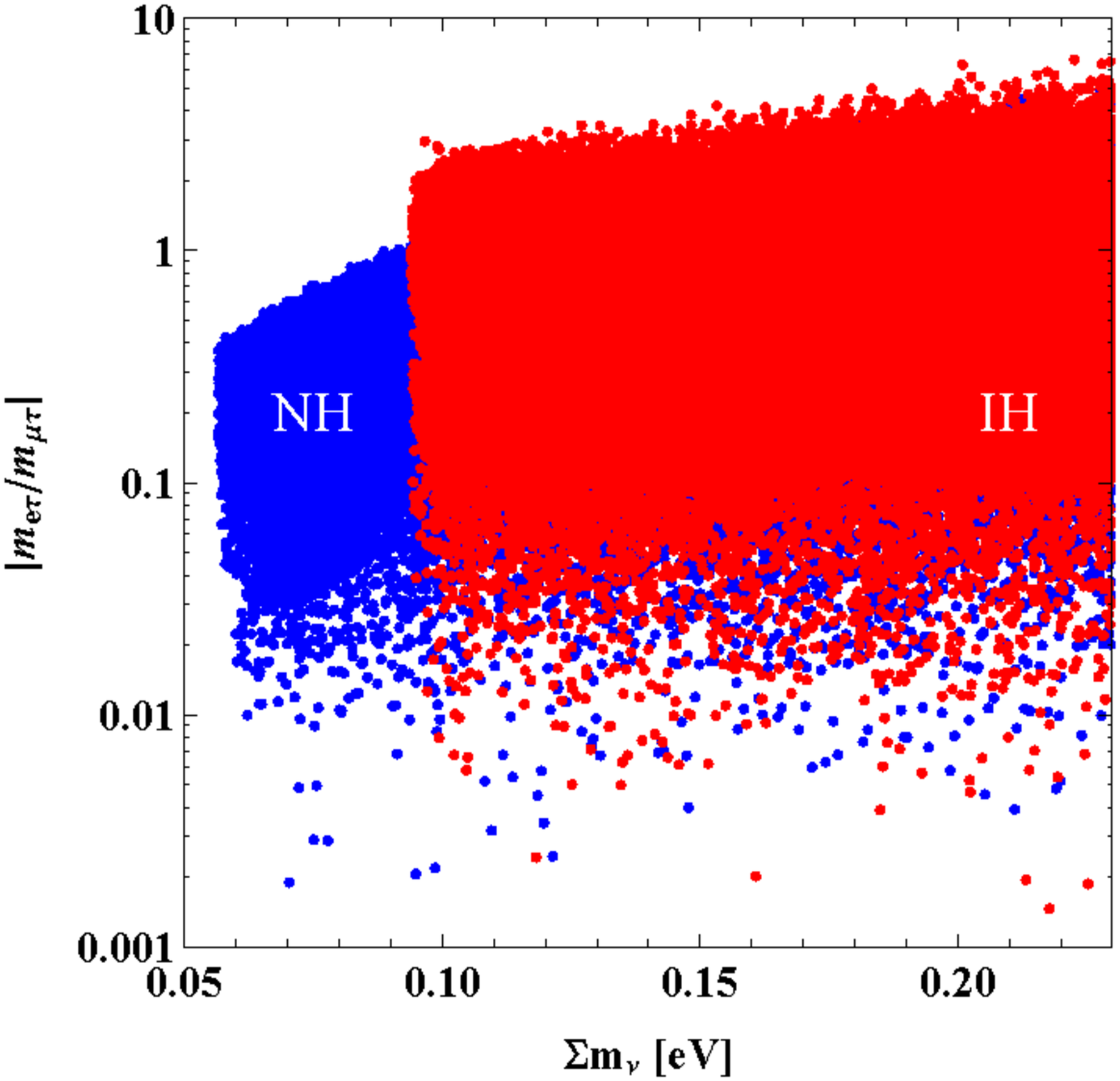}\hspace{8mm}
 \includegraphics[scale=0.293]{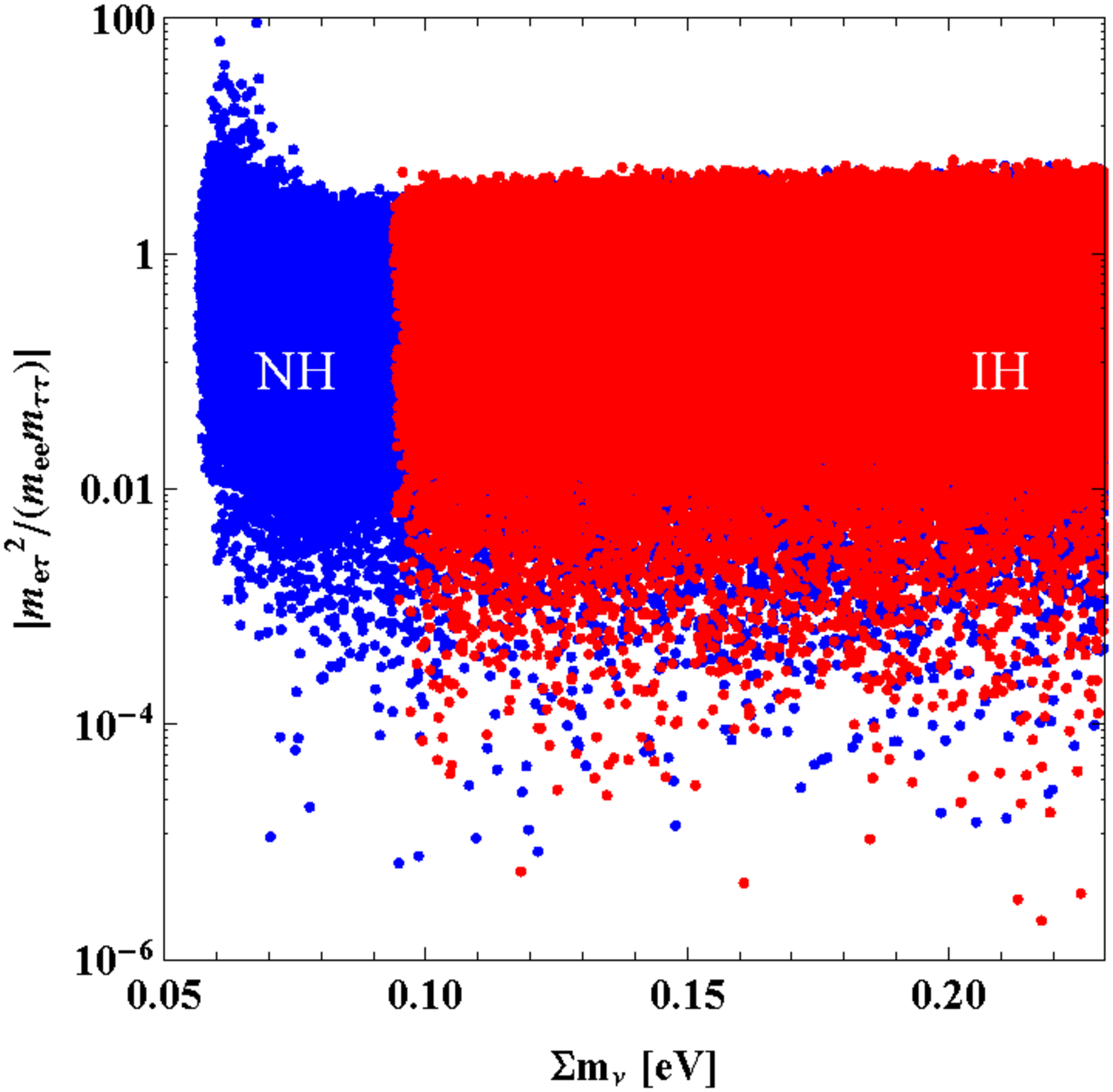}
 \end{center}
 \caption{Values of the renormalization invariants vs the sum of the neutrino 
masses in the $0\leq(\delta,\alpha,\beta)<2\pi/3$ case.}
 \label{fig1}
 \end{figure}
 \begin{figure}
 \begin{center}
 \includegraphics[scale=0.3]{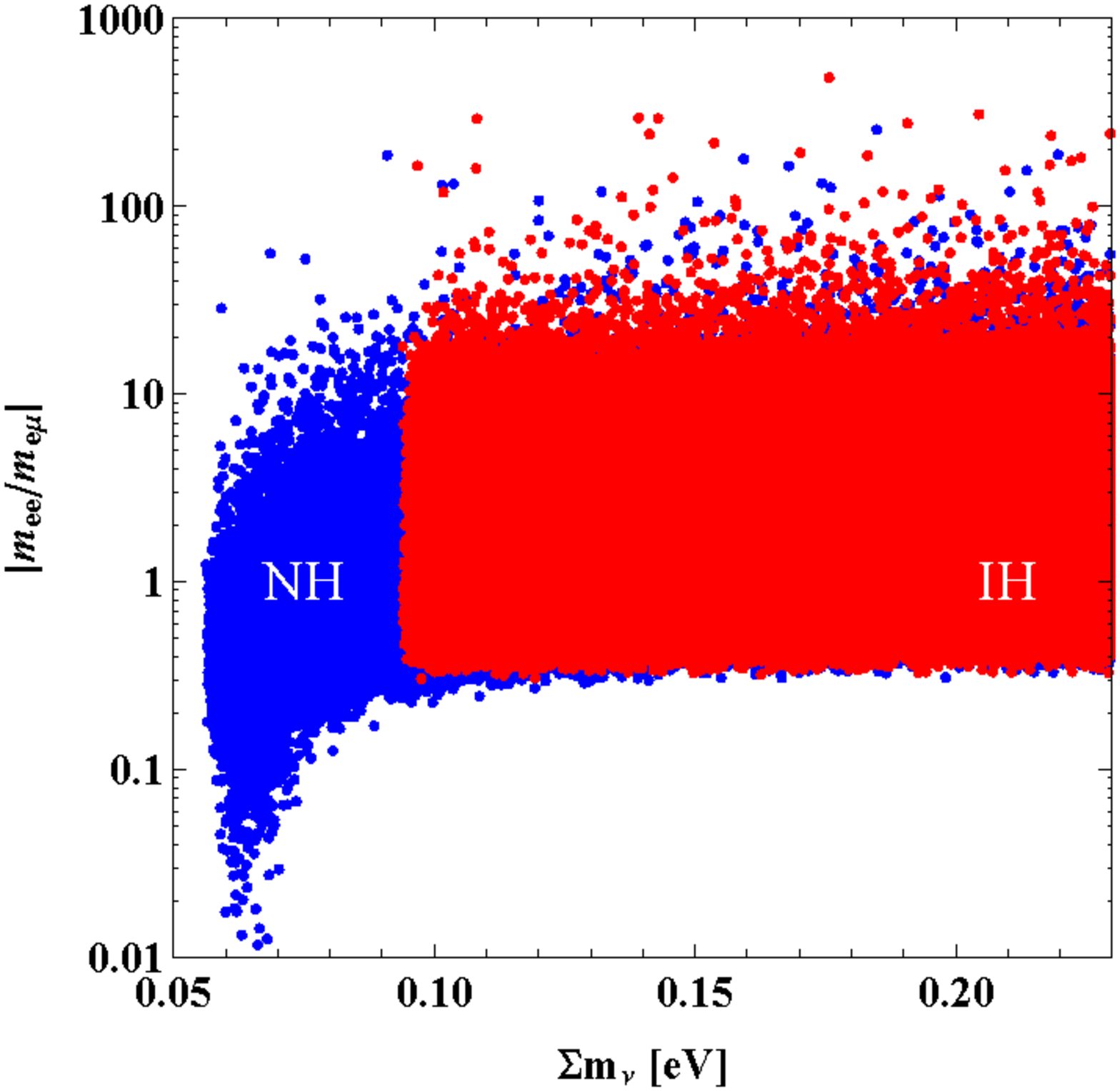}\hspace{8mm}
 \includegraphics[scale=0.3]{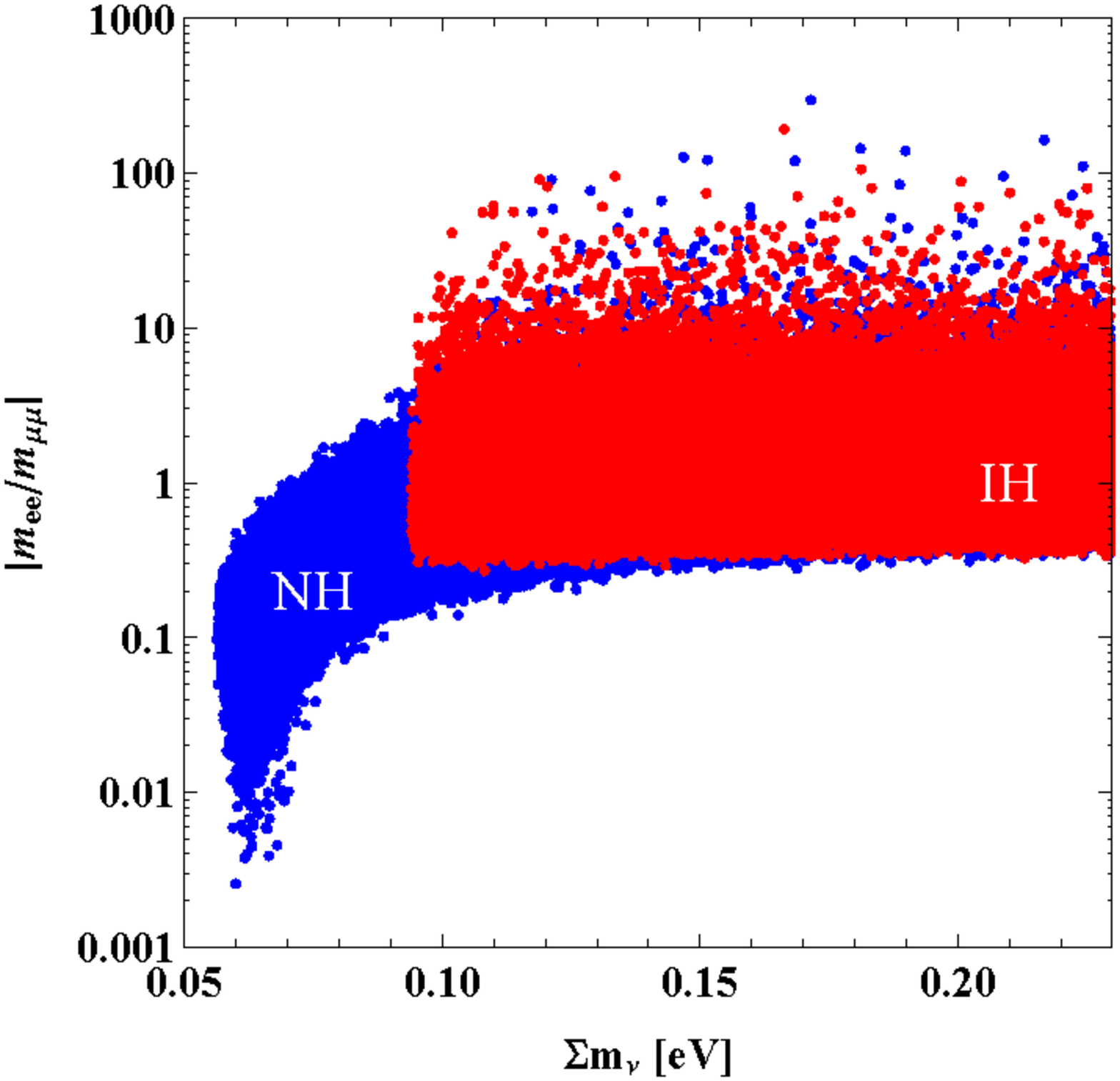}

~\\

 \includegraphics[scale=0.3]{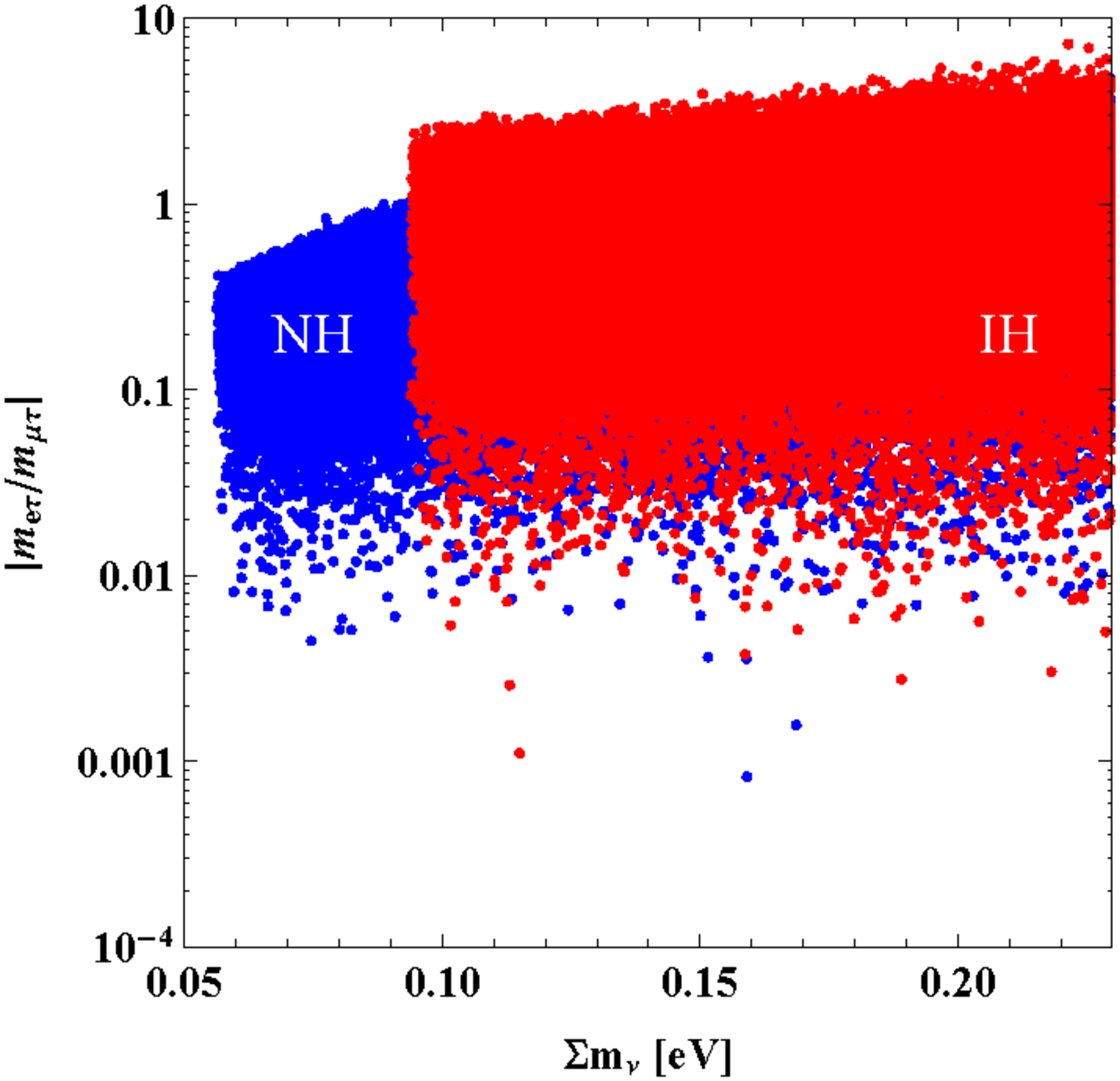}\hspace{8mm}
 \includegraphics[scale=0.294]{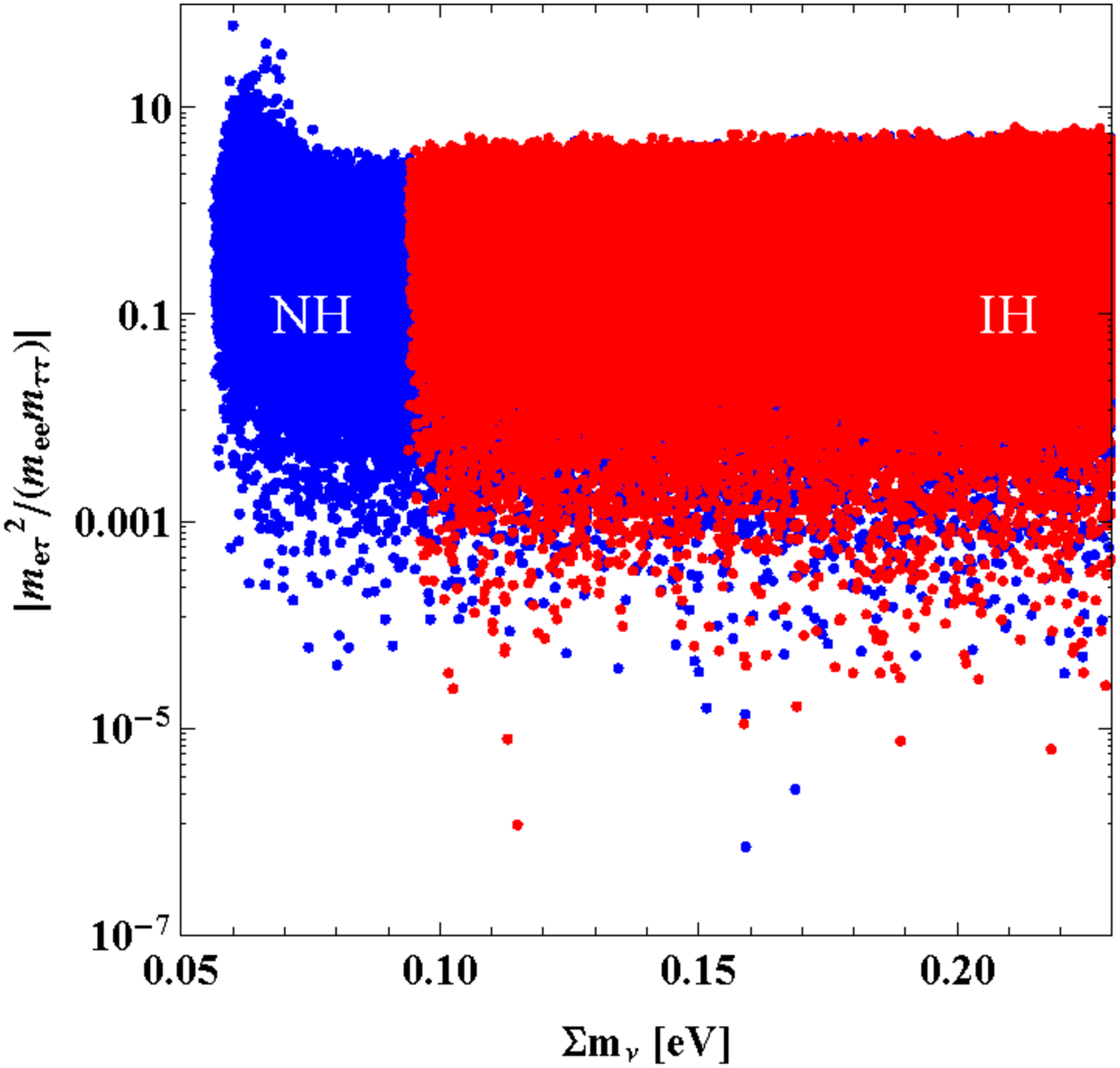}
 \end{center}
 \caption{Values of the renormalization invariants vs the sum of the neutrino 
masses in the $4\pi/3\leq(\delta,\alpha,\beta)<2\pi$.}
 \label{fig1-2}
 \end{figure}
 \begin{eqnarray}
  && 0.27\leq\sin^2\theta_{12}\leq0.34,\qquad0.34\leq\sin^2\theta_{23}\leq0.67,
     \qquad0.016\leq\sin^2\theta_{13}\leq0.030, \\
  && 7.00\times10^{-5}\mbox{ eV}^2\leq\Delta m_{21}^2\leq
     8.09\times10^{-5}\mbox{ eV}^2, \\
  && 2.27\times10^{-3}\mbox{ eV}^2\leq\Delta m_{31}^2\leq
     2.69\times10^{-3}\mbox{ eV}^2, \\
  && -2.65\times10^{-3}\mbox{ eV}^2\leq\Delta m_{32}^2\leq
     -2.24\times10^{-3}\mbox{ eV}^2.
 \end{eqnarray}
The horizontal axes in all figures of Figures~\ref{fig1} and \ref{fig1-2} are 
the sum of the three light neutrino masses $\sum m_\nu$ which is constrained by 
the first cosmological result based on {\it Planck} measurements of the cosmic 
microwave background combined with other cosmological data as $\sum m_\nu<0.23$ 
eV at 95\% CL~\cite{Ade:2013zuv}.\footnote{See also~\cite{Haba:2013xwa} for 
constraints on neutrino mass ordering and degeneracy from the {\it Planck} and 
neutrino-less double beta decay.}

We find from Figures~\ref{fig1} and \ref{fig1-2} that
 \begin{eqnarray}
  && 5\times10^{-2}\lesssim\left|\frac{m_{ee}}{m_{e\mu}}\right|\lesssim
     2\times10^2,\qquad
     10^{-3}\lesssim\left|\frac{m_{ee}}{m_{\mu\mu}}\right|\lesssim
     3\times10^2, \\
  && 2\times10^{-3}\lesssim\left|\frac{m_{e\tau}}{m_{\mu\tau}}\right|\lesssim5,
     \qquad8\times10^{-6}\lesssim\left|\frac{m_{e\tau}^2}{m_{ee}m_{\mu\tau}}
                                 \right|\lesssim10^2,
 \end{eqnarray}
for the NH with $0\leq(\delta,\alpha,\beta)<2\pi/3$, and
 \begin{eqnarray}
  && 0.3\lesssim\left|\frac{m_{ee}}{m_{e\mu}}\right|\lesssim6\times10^2,\qquad
     0.3\lesssim\left|\frac{m_{ee}}{m_{\mu\mu}}\right|\lesssim3\times10^2, \\
  && 1.5\times10^{-3}\lesssim\left|\frac{m_{e\tau}}{m_{\mu\tau}}\right|
     \lesssim7,\qquad
     2\times10^{-6}\lesssim\left|\frac{m_{e\tau}^2}{m_{ee}m_{\mu\tau}}\right|\lesssim70,
 \end{eqnarray} 
for the IH with $0\leq(\delta,\alpha,\beta)<2\pi/3$, and 
 \begin{eqnarray}
  && 10^{-2}\lesssim\left|\frac{m_{ee}}{m_{e\mu}}\right|\lesssim3\times10^2,
     \qquad3\times10^{-3}\lesssim\left|\frac{m_{ee}}{m_{\mu\mu}}\right|\lesssim
     3\times10^2, \\
  && 9\times10^{-4}\lesssim\left|\frac{m_{e\tau}}{m_{\mu\tau}}\right|\lesssim5,
     \qquad8\times10^{-7}\lesssim\left|\frac{m_{e\tau}^2}{m_{ee}m_{\mu\tau}}
                                 \right|\lesssim10^2,
 \end{eqnarray}
for the NH with $4\pi/3\leq(\delta,\alpha,\beta)<2\pi$, and
 \begin{eqnarray}
  && 0.3\lesssim\left|\frac{m_{ee}}{m_{e\mu}}\right|\lesssim5\times10^2,\qquad
     0.3\lesssim\left|\frac{m_{ee}}{m_{\mu\mu}}\right|\lesssim2\times10^2, \\
  && 10^{-3}\lesssim\left|\frac{m_{e\tau}}{m_{\mu\tau}}\right|
     \lesssim8,\qquad
     10^{-6}\lesssim\left|\frac{m_{e\tau}^2}{m_{ee}m_{\mu\tau}}\right|\lesssim6,
 \end{eqnarray} 
for the IH with $4\pi/3\leq(\delta,\alpha,\beta)<2\pi$. We can also consider 
other combination of sizes of CP phases. The results of other possible cases are
 summarized in Tables~\ref{tab1}-\ref{tab6} in Appendix.

The invariance is independent of neutrino mass ordering and a parameterization 
of the PMNS matrix. The other quantities by using these invariants are also RGE 
invariants. In a general Majorana mass matrix for the 3 generations of the light
 neutrinos, there are 9 degrees of freedom, which are described by 3 mixing 
angles, 3 masses and 3 CP-phases. Regarding the number of the RGE invariants, 3 
CP phases~\cite{Haba:1999ca} and the above 4 combinations among matrix elements 
in \eqref{hM} are the independent RGE invariants. Therefore, the remaining 2 
quantities are not RGE invariants. These are the overall factor $r$ and the 
small parameter $\epsilon$. In the next subsection, we show runnings of these 
parameters.

\subsection{Runnings of {\boldmath $r$ and $\epsilon$}}

\begin{figure}
\hspace{4.4cm}(a)\hspace{7.5cm}(b)
\begin{center}
\includegraphics[scale=0.89]{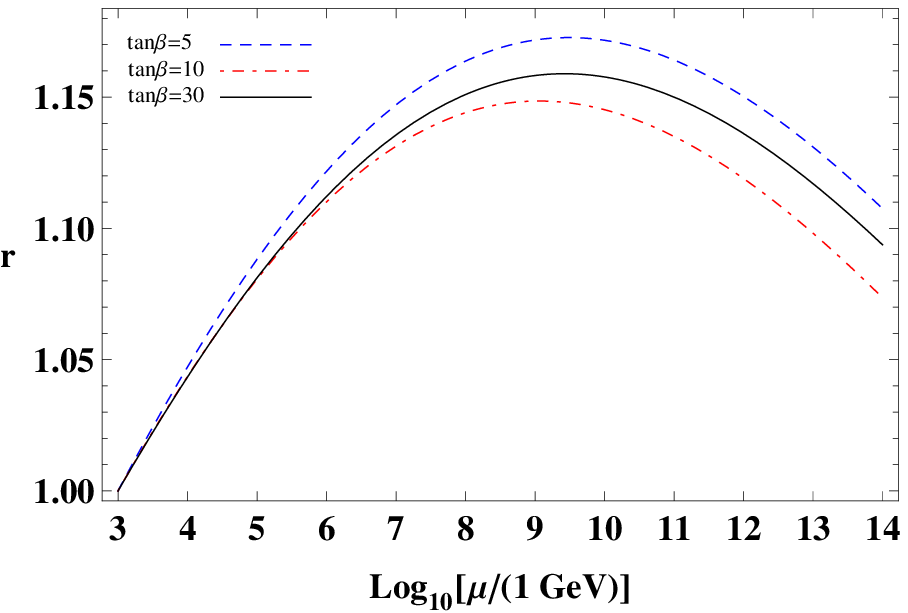}
\includegraphics[scale=0.89]{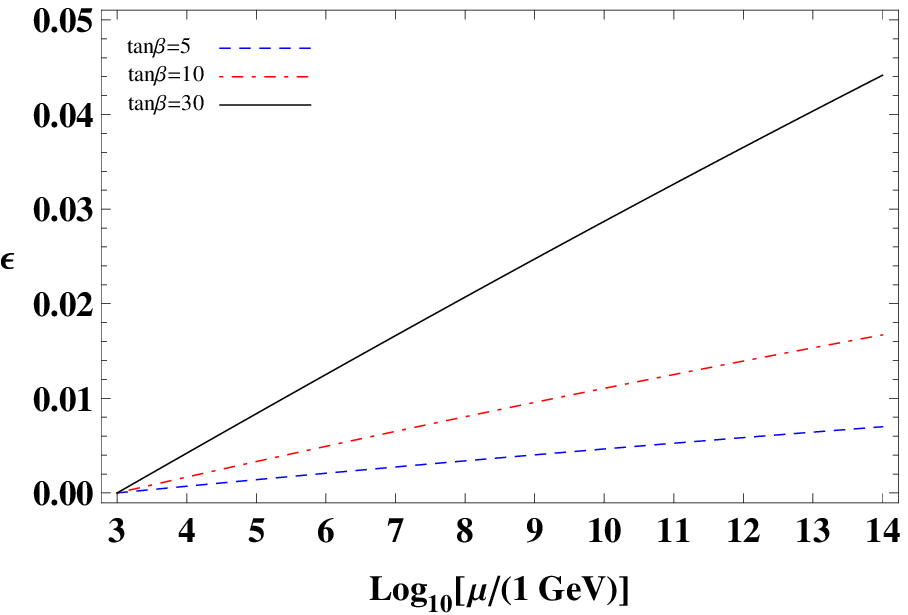}
\end{center}
\caption{Runnings of the overall factor and small parameter $\epsilon$ in the 
neutrino mass matrix.}
\label{fig2}
\end{figure}

We show runnings of $r$ and $\epsilon$ in this subsection. Clearly, the running 
of $r$ does not affects on the mixing angles while the one of $\epsilon$ affects
 on them. The running of $r$ is determined by the solving the RGE including only
 flavor blind effects as:
 \begin{eqnarray}
  16\pi^2\frac{d\kappa_{33}}{dt}
   =\bar{\alpha}\kappa_{33}
    +\left[(y_ey_e^\dagger)\kappa+\kappa(y_ey_e^\dagger)^T\right]_{33},
 \end{eqnarray}
with a relation
 \begin{eqnarray}
  r(\mu)=\frac{\kappa_{33}(\mu)}{\kappa_{33}(\Lambda_{\rm EW})(1+\epsilon)^2}
        =\frac{(M_\nu(\mu))_{\tau\tau}}
              {(M_\nu(\Lambda_{\rm EW}))_{\tau\tau}(1+\epsilon)^2}.
  \label{r}
 \end{eqnarray}
The running of $\epsilon$ is determined by RGEs for the Yukawa couplings of 
charged leptons as
 \begin{eqnarray}
  16\pi^2\frac{dy_\alpha}{dt}
  =y_\alpha\left(4y_\alpha^\ast y_\alpha+3y_b^\ast y_b-3g_2^2-\frac{9}{5}g_1^2
           \right),
 \end{eqnarray}
for $\alpha=\mu,\tau$, where $y_b$ is a Yukawa coupling of bottom quark. The 
runnings of $r$ and $\epsilon$ are given in Fig.~\ref{fig2}. In the 
calculations, a SUSY threshold is taken at $10^3$ GeV and the computations are 
shown within $10^3$ GeV$\leq\mu\leq10^{14}$ 
GeV.\footnote{We neglect running effects between the EW scale and the SUSY 
scale, for simplicity. Therefore, $r=1$ and $\epsilon=0$ are taken at $10^3$ GeV
 as boundary conditions for the RGE.} Since the neutrino Yukawa coupling exceeds
 1 at higher energy scale than $10^{14}$ GeV in the seesaw mechanism, we 
consider the energy scale lower than $10^{14}$ GeV.

\begin{figure}
\hspace{4.4cm}(a)\hspace{7.5cm}(b)
\begin{center}
\includegraphics[scale=0.89]{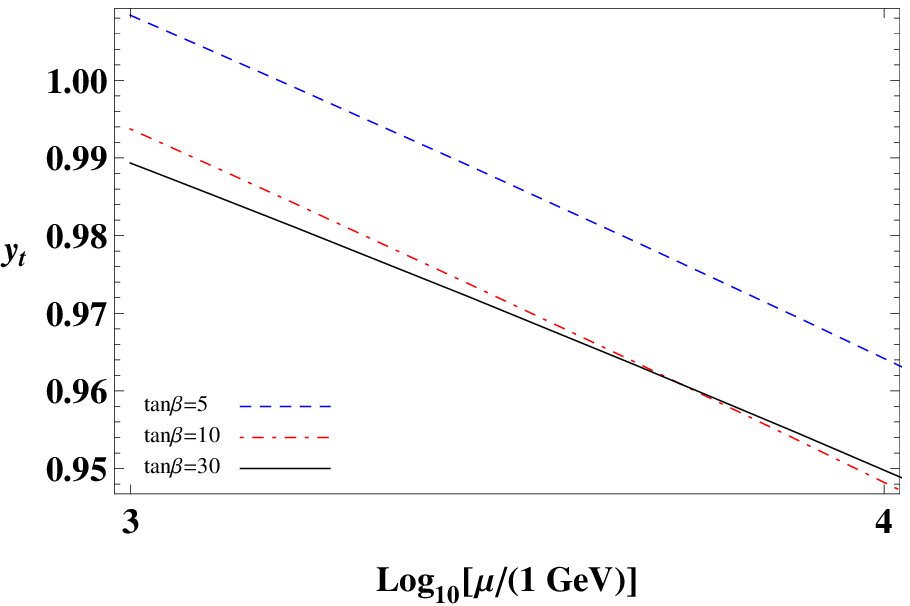}
\includegraphics[scale=0.89]{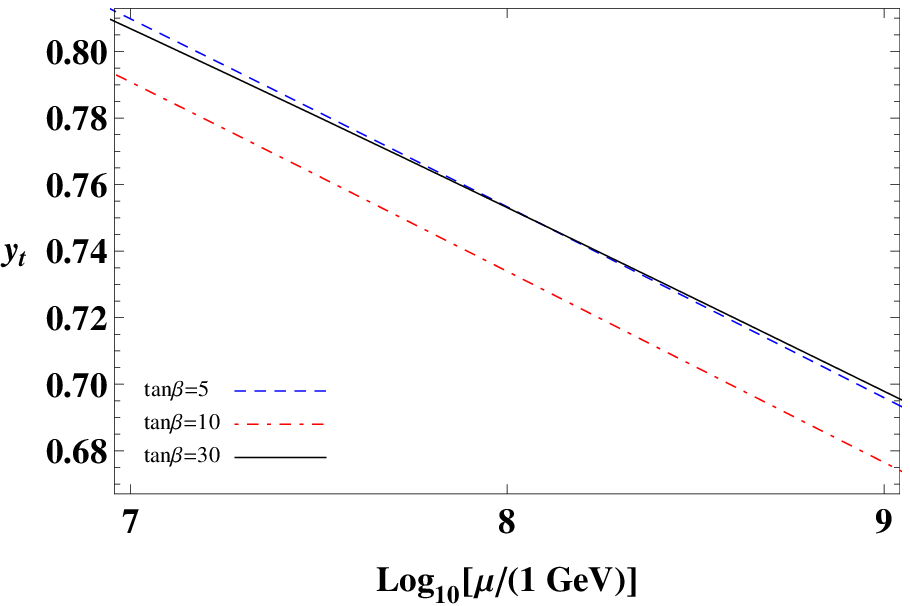}
\end{center}
\caption{Runnings of the top Yukawa coupling.}
\label{fig4}
\end{figure}

Note that these runnings are independent of the neutrino mass spectra including
 mass ordering. For the running of the overall factor $r$ in Fig.~\ref{fig2} 
(a), a contribution from the top Yukawa coupling to $\bar{\alpha}$ in \eqref{al}
 is dominant up to an energy scale lower than $\mu\sim\mathcal{O}(10^{9-10})$ 
GeV from low energy. On the other hand, contributions from the gauge couplings 
to $\bar{\alpha}$ become dominant at a higher energy scale than 
$\mu\sim\mathcal{O}(10^{9-10})$ GeV. Therefore, there are peaks around 
$\mu\sim\mathcal{O}(10^{9-10})$ GeV in the runnings of the overall factor for 
all cases of $\tan\beta$. As a result, the values of $r$ can be in ranges of 
$1.00\leq r\lesssim(1.18,1.15,1.16)$ in $\tan\beta=(5, 10, 30)$ cases for 
$10^3$ GeV$\leq\mu\leq10^{14}$ GeV, respectively. For $\tan\beta=10$ case, the 
top Yukawa coupling is smaller than that of $\tan\beta=5$ case up to a high 
energy scale as $\mu\simeq10^{14}$ GeV. Thus, the curve of running 
for $\tan\beta=10$ case is lower than one of $\tan\beta=5$ case and the position
 of the peak appears at smaller $r$ and lower $\mu$ compared to $\tan\beta=5$ 
case. In particular, the position of the peak for $\tan\beta=10$ case appears at
 the smallest $r$ and $\mu$ among all cases. On the other hand, the top Yukawa 
coupling for $\tan\beta=30$ case is the smallest among all cases of $\tan\beta$ 
at $10^3$ GeV as shown in Fig.~\ref{fig4} (a), but it becomes the largest at 
higher energy scale because of a non-negligible contribution from a large bottom
 Yukawa coupling to a running of the top Yukawa as shown in Fig.~\ref{fig4} (b).
The values of $r=1$ return at $\mu\simeq2.20\times10^{17},~1.11\times10^{16}$, 
and $2.82\times10^{15}$ GeV in $\tan\beta=5$, 10, and 30 cases, respectively. In
 the cases, values of the matrix elements such as $(M_\nu(\mu))_{ee}$, 
$(M_\nu(\mu))_{e\mu}$, and $(M_\nu(\mu))_{\mu\mu}$ are the same as ones at the 
low energy scale. However, the neutrino Yukawa coupling
 exceeds 1 at these energy scales so that we do not consider anymore $r=1$ 
returning point at the high energy. 
As far as the running of small parameter $\epsilon$ concerned, the values are 
smaller than $7.00\times10^{-3}$, $1.39\times10^{-2}$, and  $4.42\times10^{-2}$ 
for $\tan\beta=5$, 10, and 30, respectively when one considers the 
renormalization scale up to $10^{14}$ GeV. By using the values of $r$ and 
$\epsilon$ shown in Fig.~\ref{fig2} with experimentally observed values of the 
neutrino parameters, one can determine a values of matrix elements in \eqref{hM}
 at an arbitrary high energy scale.

\section{Summary}

We investigated a behavior of the coefficient of the Weinberg operator, which 
describes tiny neutrino masses, under the RGEs at arbitrary high energy regime 
because the neutrinos might be a key for a study of new physics beyond the SM. A
 simple analytical discussion clarified that there are several renormalization 
invariants, which are described by ratios among elements of the Majorana mass 
matrix of the light neutrinos as $m_{ee}/m_{e\mu}$, $m_{ee}/m_{\mu\mu}$, 
$m_{e\tau}/m_{\mu\tau}$, and $m_{e\tau}^2/(m_{ee}m_{\tau\tau})$. The invariance 
is independent of neutrino mass ordering and a parameterization of the PMNS 
matrix. The values are within the ranges of 
$10^{-3}\lesssim|m_{ee}/m_{e\mu}|\lesssim3\times10^3$, 
$10^{-3}\lesssim|m_{ee}/m_{\mu\mu}|\lesssim6\times10^2$, 
$1.2\times10^{-4}\lesssim|m_{e\tau}/m_{\mu\tau}|\lesssim6$, and 
$8\times10^{-7}\lesssim|m_{e\tau}^2/(m_{ee}m_{\mu\tau})|\lesssim10^2$ for the NH
 and $0.2\lesssim|m_{ee}/m_{e\mu}|\lesssim1.8\times10^3$, 
$0.2\lesssim|m_{ee}/m_{\mu\mu}|\lesssim8\times10^2$, 
$10^{-3}\lesssim|m_{e\tau}/m_{\mu\tau}|\lesssim8$, and 
$8\times10^{-7}\lesssim|m_{e\tau}^2/(m_{ee}m_{\mu\tau})|\lesssim70$ for the IH.

Next we estimated the runnings of the overall factor $r$ and small parameter 
$\epsilon$ under the corresponding RGEs. Once one determines these parameters at
 arbitrary high energy scale, one can obtain the values of neutrino mass 
matrix elements at the high energy. As a result, we found that the values of $r$
 can be in ranges of 
$1\leq r\lesssim(1.18,1.15,1.16)$ in $\tan\beta=(5, 10, 30)$ cases for 
$10^3$ GeV$\leq\mu\leq10^{14}$ GeV, respectively. 
As far as the running of small parameter $\epsilon$ concerned, the values are 
smaller than $7.00\times10^{-3}$, $1.39\times10^{-2}$, and  $4.42\times10^{-2}$ 
for $\tan\beta=5$, 10, and 30, respectively when one considers the 
renormalization scale up to $10^{14}$ GeV.

\subsection*{Acknowledgement}

This work is partially supported by Scientific Grant by Ministry of Education 
and Science, Nos. 00293803, 20244028, 21244036, 23340070, and by the SUHARA Memorial Foundation. The work of R.T. 
is supported by Research Fellowships of the Japan Society for the Promotion of 
Science for Young Scientists.

\appendix
\section*{Appendix}

We show the numerical results of renormalization group invariants for other possible combinations of the CP phases in the following Tables~\ref{tab1}-\ref{tab6}.

\begin{table}
\begin{center}
\begin{tabular}{|c|c|}
\hline
\multicolumn{2}{|c|}{$0\leq(\delta,\alpha)<2\pi/3$ and $2\pi/3\leq\beta<4\pi/3$}\\
\hline\hline
$3\times10^{-3}\lesssim\left|m_{ee}/m_{e\mu}\right|\lesssim3\times10^2$ & $1.5\times10^{-3}\lesssim\left|m_{ee}/m_{\mu\mu}\right|\lesssim2\times10^2$  \\ 
\hline
$2\times10^{-3}\lesssim\left|m_{e\tau}/m_{\mu\tau}\right|\lesssim6$ & $10^{-5}\lesssim\left|m_{e\tau}^2/(m_{ee}m_{\mu\tau})\right|\lesssim10^2$ \\
\hline
\multicolumn{2}{c}{~}\\
\hline
\multicolumn{2}{|c|}{$0\leq(\delta,\alpha)<2\pi/3$ and $4\pi/3\leq\beta<2\pi$}\\
\hline\hline
$2\times10^{-3}\lesssim\left|m_{ee}/m_{e\mu}\right|\lesssim3\times10^2$ & $10^{-3}\lesssim\left|m_{ee}/m_{\mu\mu}\right|\lesssim1.5\times10^2$  \\ 
\hline
$5\times10^{-4}\lesssim\left|m_{e\tau}/m_{\mu\tau}\right|\lesssim4$ & $8\times10^{-7}\lesssim\left|m_{e\tau}^2/(m_{ee}m_{\mu\tau})\right|\lesssim50$ \\
\hline
\multicolumn{2}{c}{~}\\
\hline
\multicolumn{2}{|c|}{$0\leq(\delta,\beta)<2\pi/3$ and $2\pi/3\leq\alpha<4\pi/3$}\\
\hline\hline
$0.1\lesssim\left|m_{ee}/m_{e\mu}\right|\lesssim5\times10^2$ & $2\times10^{-2}\lesssim\left|m_{ee}/m_{\mu\mu}\right|\lesssim3\times10^2$  \\ 
\hline
$6\times10^{-4}\lesssim\left|m_{e\tau}/m_{\mu\tau}\right|\lesssim5$ & $10^{-6}\lesssim\left|m_{e\tau}^2/(m_{ee}m_{\mu\tau})\right|\lesssim5$ \\
\hline
\multicolumn{2}{c}{~}\\
\hline
\multicolumn{2}{|c|}{$0\leq\delta<2\pi/3$ and $2\pi/3\leq(\alpha,\beta)<4\pi/3$}\\
\hline\hline
$0.1\lesssim\left|m_{ee}/m_{e\mu}\right|\lesssim4\times10^2$ & $3\times10^{-2}\lesssim\left|m_{ee}/m_{\mu\mu}\right|\lesssim1.1\times10^2$  \\ 
\hline
$1.2\times10^{-4}\lesssim\left|m_{e\tau}/m_{\mu\tau}\right|\lesssim6$ & $3\times10^{-8}\lesssim\left|m_{e\tau}^2/(m_{ee}m_{\mu\tau})\right|\lesssim5$  \\
\hline
\multicolumn{2}{c}{~}\\
\hline
\multicolumn{2}{|c|}{$0\leq\delta<2\pi/3$, $2\pi/3\leq\alpha<4\pi/3$ and $4\pi/3\leq\beta<2\pi$}\\
\hline\hline
$8\times10^{-2}\lesssim\left|m_{ee}/m_{e\mu}\right|\lesssim3\times10^3$ & $2\times10^{-2}\lesssim\left|m_{ee}/m_{\mu\mu}\right|\lesssim1.5\times10^2$ \\ 
\hline
$5\times10^{-4}\lesssim\left|m_{e\tau}/m_{\mu\tau}\right|\lesssim3.5$ & $10^{-6}\lesssim\left|m_{e\tau}^2/(m_{ee}m_{\mu\tau})\right|\lesssim6$ \\
\hline
\multicolumn{2}{c}{~}\\
\hline
\multicolumn{2}{|c|}{$0\leq(\delta,\beta)<2\pi/3$ and $4\pi/3\leq\alpha<2\pi$}\\
\hline\hline
$4\times10^{-3}\lesssim\left|m_{ee}/m_{e\mu}\right|\lesssim9\times10^2$ & $1.2\times10^{-3}\lesssim\left|m_{ee}/m_{\mu\mu}\right|\lesssim2\times10^2$  \\ 
\hline
$2\times10^{-3}\lesssim\left|m_{e\tau}/m_{\mu\tau}\right|\lesssim3.5$ & $10^{-5}\lesssim\left|m_{e\tau}^2/(m_{ee}m_{\mu\tau})\right|\lesssim10^2$ \\
\hline
\multicolumn{2}{c}{~}\\
\hline
\multicolumn{2}{|c|}{$0\leq\delta<2\pi/3$, $4\pi/3\leq\alpha<2\pi$ and $2\pi/3\leq\beta<4\pi/3$}\\
\hline\hline
$4\times10^{-3}\lesssim\left|m_{ee}/m_{e\mu}\right|\lesssim2.5\times10^2$ & $1.2\times10^{-3}\lesssim\left|m_{ee}/m_{\mu\mu}\right|\lesssim2\times10^2$  \\ 
\hline
$2\times10^{-3}\lesssim\left|m_{e\tau}/m_{\mu\tau}\right|\lesssim3.2$ & $10^{-5}\lesssim\left|m_{e\tau}^2/(m_{ee}m_{\mu\tau})\right|\lesssim80$ \\
\hline
\multicolumn{2}{c}{~}\\
\hline
\multicolumn{2}{|c|}{$0\leq\delta<2\pi/3$ and $4\pi/3\leq(\alpha,\beta)<2\pi$}\\
\hline\hline
$6\times10^{-3}\lesssim\left|m_{ee}/m_{e\mu}\right|\lesssim7\times10^2$ & $2\times10^{-3}\lesssim\left|m_{ee}/m_{\mu\mu}\right|\lesssim2\times10^2$  \\ 
\hline
$6\times10^{-4}\lesssim\left|m_{e\tau}/m_{\mu\tau}\right|\lesssim3$ & $5\times10^{-5}\lesssim\left|m_{e\tau}^2/(m_{ee}m_{\mu\tau})\right|\lesssim40$ \\
\hline
\end{tabular}
\end{center}
\caption{Values of the renormalization invariants for the NH.}
\label{tab1}
\end{table}
\begin{table}
\begin{center}
\begin{tabular}{|c|c|}
\hline
\multicolumn{2}{|c|}{$2\pi/3\leq\delta<4\pi/3$ and $0\leq(\alpha,\beta)<2\pi/3$}\\
\hline\hline
$8\times10^{-3}\lesssim\left|m_{ee}/m_{e\mu}\right|\lesssim1.3\times10^3$ & $2\times10^{-3}\lesssim\left|m_{ee}/m_{\mu\mu}\right|\lesssim3\times10^2$  \\ 
\hline
$1.8\times10^{-3}\lesssim\left|m_{e\tau}/m_{\mu\tau}\right|\lesssim4.6$ & $9\times10^{-5}\lesssim\left|m_{e\tau}^2/(m_{ee}m_{\mu\tau})\right|\lesssim70$ \\
\hline
\multicolumn{2}{c}{~}\\
\hline
\hline
\multicolumn{2}{|c|}{$2\pi/3\leq(\delta,\beta)<4\pi/3$ and $0\leq\alpha<2\pi/3$}\\
\hline\hline
$10^{-3}\lesssim\left|m_{ee}/m_{e\mu}\right|\lesssim7\times10^2$ & $2\times10^{-3}\lesssim\left|m_{ee}/m_{\mu\mu}\right|\lesssim3\times10^2$  \\ 
\hline
$1.3\times10^{-3}\lesssim\left|m_{e\tau}/m_{\mu\tau}\right|\lesssim4.5$ & $1.5\times10^{-6}\lesssim\left|m_{e\tau}^2/(m_{ee}m_{\mu\tau})\right|\lesssim10^2$ \\
\hline
\multicolumn{2}{c}{~}\\
\hline
\multicolumn{2}{|c|}{$2\pi/3\leq\delta<4\pi/3$, $0\leq\alpha<2\pi/3$ and $4\pi/3\leq\beta<2\pi$}\\
\hline\hline
$9\times10^{-3}\lesssim\left|m_{ee}/m_{e\mu}\right|\lesssim1.5\times10^2$ & $2\times10^{-3}\lesssim\left|m_{ee}/m_{\mu\mu}\right|\lesssim2.5\times10^2$  \\ 
\hline
$2.8\times10^{-3}\lesssim\left|m_{e\tau}/m_{\mu\tau}\right|\lesssim3$ & $10^{-5}\lesssim\left|m_{e\tau}^2/(m_{ee}m_{\mu\tau})\right|\lesssim50$ \\
\hline
\multicolumn{2}{c}{~}\\
\hline
\multicolumn{2}{|c|}{$2\pi/3\leq(\delta,\alpha)<4\pi/3$ and $0\leq\beta<2\pi/3$}\\
\hline\hline
$0.1\lesssim\left|m_{ee}/m_{e\mu}\right|\lesssim5\times10^2$ & $4\times10^{-2}\lesssim\left|m_{ee}/m_{\mu\mu}\right|\lesssim6\times10^2$  \\ 
\hline
$2.5\times10^{-3}\lesssim\left|m_{e\tau}/m_{\mu\tau}\right|\lesssim5$ & $5\times10^{-6}\lesssim\left|m_{e\tau}^2/(m_{ee}m_{\mu\tau})\right|\lesssim5$ \\
\hline
\multicolumn{2}{c}{~}\\
\hline
\multicolumn{2}{|c|}{$2\pi/3\leq(\delta,\alpha,\beta)<4\pi/3$}\\
\hline\hline
$0.1\lesssim\left|m_{ee}/m_{e\mu}\right|\lesssim6\times10^2$ & $3.5\times10^{-2}\lesssim\left|m_{ee}/m_{\mu\mu}\right|\lesssim2.5\times10^2$  \\ 
\hline
$2\times10^{-3}\lesssim\left|m_{e\tau}/m_{\mu\tau}\right|\lesssim5$ & $5\times10^{-6}\lesssim\left|m_{e\tau}^2/(m_{ee}m_{\mu\tau})\right|\lesssim5$  \\
\hline
\multicolumn{2}{c}{~}\\
\hline
\multicolumn{2}{|c|}{$2\pi/3\leq(\delta,\alpha)<4\pi/3$ and $4\pi/3\leq\beta<2\pi$}\\
\hline\hline
$0.1\lesssim\left|m_{ee}/m_{e\mu}\right|\lesssim6\times10^2$ & $3\times10^{-2}\lesssim\left|m_{ee}/m_{\mu\mu}\right|\lesssim4\times10^2$ \\ 
\hline
$3\times10^{-3}\lesssim\left|m_{e\tau}/m_{\mu\tau}\right|\lesssim4.5$ & $10^{-5}\lesssim\left|m_{e\tau}^2/(m_{ee}m_{\mu\tau})\right|\lesssim5$ \\
\hline
\multicolumn{2}{c}{~}\\
\hline
\multicolumn{2}{|c|}{$2\pi/3\leq\delta<4\pi/3$, $4\pi/3\leq\alpha<2\pi$ and $0\leq\beta<2\pi/3$}\\
\hline\hline
$4\times10^{-3}\lesssim\left|m_{ee}/m_{e\mu}\right|\lesssim4\times10^2$ & $4\times10^{-3}\lesssim\left|m_{ee}/m_{\mu\mu}\right|\lesssim2\times10^2$  \\ 
\hline
$2\times10^{-3}\lesssim\left|m_{e\tau}/m_{\mu\tau}\right|\lesssim3$ & $3\times10^{-6}\lesssim\left|m_{e\tau}^2/(m_{ee}m_{\mu\tau})\right|\lesssim40$ \\
\hline
\multicolumn{2}{c}{~}\\
\hline
\multicolumn{2}{|c|}{$2\pi/3\leq(\delta,\beta)<4\pi/3$ and $4\pi/3\leq\alpha<2\pi$}\\
\hline\hline
$7\times10^{-3}\lesssim\left|m_{ee}/m_{e\mu}\right|\lesssim10^3$ & $2\times10^{-3}\lesssim\left|m_{ee}/m_{\mu\mu}\right|\lesssim10^2$  \\ 
\hline
$2\times10^{-3}\lesssim\left|m_{e\tau}/m_{\mu\tau}\right|\lesssim4.5$ & $3\times10^{-6}\lesssim\left|m_{e\tau}^2/(m_{ee}m_{\mu\tau})\right|\lesssim30$ \\
\hline
\multicolumn{2}{c}{~}\\
\hline
\multicolumn{2}{|c|}{$2\pi/3\leq\delta<4\pi/3$ and $4\pi/3\leq(\alpha,\beta)<2\pi$}\\
\hline\hline
$10^{-2}\lesssim\left|m_{ee}/m_{e\mu}\right|\lesssim9\times10^2$ & $2\times10^{-3}\lesssim\left|m_{ee}/m_{\mu\mu}\right|\lesssim1.5\times10^2$  \\ 
\hline
$1.3\times10^{-3}\lesssim\left|m_{e\tau}/m_{\mu\tau}\right|\lesssim4.5$ & $10^{-5}\lesssim\left|m_{e\tau}^2/(m_{ee}m_{\mu\tau})\right|\lesssim10^2$ \\
\hline
\end{tabular}
\end{center}
\caption{Values of the renormalization invariants for the NH.}
\label{tab2}
\end{table}
\begin{table}
\begin{center}
\begin{tabular}{|c|c|}
\hline
\multicolumn{2}{|c|}{$4\pi/3\leq\delta<2\pi$ and $0\leq(\alpha,\beta)<2\pi/3$}\\
\hline\hline
$2\times10^{-3}\lesssim\left|m_{ee}/m_{e\mu}\right|\lesssim8\times10^2$ & $2.5\times10^{-3}\lesssim\left|m_{ee}/m_{\mu\mu}\right|\lesssim2\times10^2$  \\ 
\hline
$1.3\times10^{-3}\lesssim\left|m_{e\tau}/m_{\mu\tau}\right|\lesssim3.2$ & $1.5\times10^{-5}\lesssim\left|m_{e\tau}^2/(m_{ee}m_{\mu\tau})\right|\lesssim10^2$ \\
\hline
\multicolumn{2}{c}{~}\\
\hline
\hline
\multicolumn{2}{|c|}{$4\pi/3\leq\delta<2\pi$, $0\leq\alpha<2\pi/3$ and $2\pi/3\leq\beta<4\pi/3$}\\
\hline\hline
$9\times10^{-3}\lesssim\left|m_{ee}/m_{e\mu}\right|\lesssim4\times10^2$ & $3\times10^{-3}\lesssim\left|m_{ee}/m_{\mu\mu}\right|\lesssim50$  \\ 
\hline
$2.8\times10^{-3}\lesssim\left|m_{e\tau}/m_{\mu\tau}\right|\lesssim3.5$ & $1.5\times10^{-5}\lesssim\left|m_{e\tau}^2/(m_{ee}m_{\mu\tau})\right|\lesssim60$ \\
\hline
\multicolumn{2}{c}{~}\\
\hline
\multicolumn{2}{|c|}{$4\pi/3\leq(\delta,\beta)<2\pi$ and $0\leq\alpha<2\pi/3$}\\
\hline\hline
$9\times10^{-3}\lesssim\left|m_{ee}/m_{e\mu}\right|\lesssim3.5\times10^2$ & $2.5\times10^{-3}\lesssim\left|m_{ee}/m_{\mu\mu}\right|\lesssim5\times10^2$  \\ 
\hline
$4\times10^{-3}\lesssim\left|m_{e\tau}/m_{\mu\tau}\right|\lesssim3.5$ & $1.2\times10^{-4}\lesssim\left|m_{e\tau}^2/(m_{ee}m_{\mu\tau})\right|\lesssim50$ \\
\hline
\multicolumn{2}{c}{~}\\
\hline
\multicolumn{2}{|c|}{$4\pi/3\leq\delta<2\pi$, $2\pi/3\leq\alpha<4\pi/3$ and $0\leq\beta<2\pi/3$}\\
\hline\hline
$0.1\lesssim\left|m_{ee}/m_{e\mu}\right|\lesssim4\times10^2$ & $4\times10^{-2}\lesssim\left|m_{ee}/m_{\mu\mu}\right|\lesssim2\times10^2$  \\ 
\hline
$1.5\times10^{-3}\lesssim\left|m_{e\tau}/m_{\mu\tau}\right|\lesssim3.2$ & $5\times10^{-6}\lesssim\left|m_{e\tau}^2/(m_{ee}m_{\mu\tau})\right|\lesssim5$ \\
\hline
\multicolumn{2}{c}{~}\\
\hline
\multicolumn{2}{|c|}{$4\pi/3\leq\delta<2\pi$ and $2\pi/3\leq(\alpha,\beta)<4\pi/3$}\\
\hline\hline
$0.1\lesssim\left|m_{ee}/m_{e\mu}\right|\lesssim7\times10^2$ & $3\times10^{-2}\lesssim\left|m_{ee}/m_{\mu\mu}\right|\lesssim10^2$  \\ 
\hline
$1.2\times10^{-3}\lesssim\left|m_{e\tau}/m_{\mu\tau}\right|\lesssim5.5$ & $5\times10^{-6}\lesssim\left|m_{e\tau}^2/(m_{ee}m_{\mu\tau})\right|\lesssim5$  \\
\hline
\multicolumn{2}{c}{~}\\
\hline
\multicolumn{2}{|c|}{$4\pi/3\leq(\delta,\beta)<2\pi$ and $2\pi/3\leq\alpha<4\pi/3$}\\
\hline\hline
$0.12\lesssim\left|m_{ee}/m_{e\mu}\right|\lesssim6\times10^2$ & $3\times10^{-2}\lesssim\left|m_{ee}/m_{\mu\mu}\right|\lesssim5\times10^2$ \\ 
\hline
$1.2\times10^{-3}\lesssim\left|m_{e\tau}/m_{\mu\tau}\right|\lesssim5$ & $2\times10^{-6}\lesssim\left|m_{e\tau}^2/(m_{ee}m_{\mu\tau})\right|\lesssim5$ \\
\hline
\multicolumn{2}{c}{~}\\
\hline
\multicolumn{2}{|c|}{$4\pi/3\leq(\delta,\alpha)<2\pi$ and $0\leq\beta<2\pi/3$}\\
\hline\hline
$5\times10^{-3}\lesssim\left|m_{ee}/m_{e\mu}\right|\lesssim4\times10^2$ & $2\times10^{-3}\lesssim\left|m_{ee}/m_{\mu\mu}\right|\lesssim1.5\times10^2$  \\ 
\hline
$1.2\times10^{-3}\lesssim\left|m_{e\tau}/m_{\mu\tau}\right|\lesssim3.5$ & $3\times10^{-6}\lesssim\left|m_{e\tau}^2/(m_{ee}m_{\mu\tau})\right|\lesssim30$ \\
\hline
\multicolumn{2}{c}{~}\\
\hline
\multicolumn{2}{|c|}{$4\pi/3\leq(\delta,\alpha)<2\pi$ and $2\pi/3\leq\beta<4\pi/3$}\\
\hline\hline
$6\times10^{-3}\lesssim\left|m_{ee}/m_{e\mu}\right|\lesssim6\times10^2$ & $2.5\times10^{-3}\lesssim\left|m_{ee}/m_{\mu\mu}\right|\lesssim10^2$  \\ 
\hline
$2\times10^{-3}\lesssim\left|m_{e\tau}/m_{\mu\tau}\right|\lesssim4.5$ & $3\times10^{-6}\lesssim\left|m_{e\tau}^2/(m_{ee}m_{\mu\tau})\right|\lesssim40$ \\
\hline
\end{tabular}
\end{center}
\caption{Values of the renormalization invariants for the NH.}
\label{tab3}
\end{table}
\begin{table}
\begin{center}
\begin{tabular}{|c|c|}
\hline
\multicolumn{2}{|c|}{$0\leq(\delta,\alpha)<2\pi/3$ and $2\pi/3\leq\beta<4\pi/3$}\\
\hline\hline
$0.3\lesssim\left|m_{ee}/m_{e\mu}\right|\lesssim5\times10^2$ & $0.3\lesssim\left|m_{ee}/m_{\mu\mu}\right|\lesssim7\times10^2$  \\ 
\hline
$1.8\times10^{-3}\lesssim\left|m_{e\tau}/m_{\mu\tau}\right|\lesssim6.5$ & $1.5\times10^{-6}\lesssim\left|m_{e\tau}^2/(m_{ee}m_{\mu\tau})\right|\lesssim3$ \\
\hline
\multicolumn{2}{c}{~}\\
\hline
\multicolumn{2}{|c|}{$0\leq(\delta,\alpha)<2\pi/3$ and $4\pi/3\leq\beta<2\pi$}\\
\hline\hline
$0.3\lesssim\left|m_{ee}/m_{e\mu}\right|\lesssim7\times10^2$ & $0.3\lesssim\left|m_{ee}/m_{\mu\mu}\right|\lesssim4\times10^2$  \\ 
\hline
$1.8\times10^{-3}\lesssim\left|m_{e\tau}/m_{\mu\tau}\right|\lesssim3.2$ & $2\times10^{-6}\lesssim\left|m_{e\tau}^2/(m_{ee}m_{\mu\tau})\right|\lesssim0.6$ \\
\hline
\multicolumn{2}{c}{~}\\
\hline
\multicolumn{2}{|c|}{$0\leq(\delta,\beta)<2\pi/3$ and $2\pi/3\leq\alpha<4\pi/3$}\\
\hline\hline
$0.7\lesssim\left|m_{ee}/m_{e\mu}\right|\lesssim1.2\times10^3$ & $0.5\lesssim\left|m_{ee}/m_{\mu\mu}\right|\lesssim4\times10^2$  \\ 
\hline
$3\times10^{-3}\lesssim\left|m_{e\tau}/m_{\mu\tau}\right|\lesssim8$ & $8\times10^{-6}\lesssim\left|m_{e\tau}^2/(m_{ee}m_{\mu\tau})\right|\lesssim5$ \\
\hline
\multicolumn{2}{c}{~}\\
\hline
\multicolumn{2}{|c|}{$0\leq\delta<2\pi/3$ and $2\pi/3\leq(\alpha,\beta)<4\pi/3$}\\
\hline\hline
$0.7\lesssim\left|m_{ee}/m_{e\mu}\right|\lesssim1.7\times10^3$ & $0.5\lesssim\left|m_{ee}/m_{\mu\mu}\right|\lesssim3\times10^2$  \\ 
\hline
$1.2\times10^{-3}\lesssim\left|m_{e\tau}/m_{\mu\tau}\right|\lesssim8$ & $8\times10^{-7}\lesssim\left|m_{e\tau}^2/(m_{ee}m_{\mu\tau})\right|\lesssim5$  \\
\hline
\multicolumn{2}{c}{~}\\
\hline
\multicolumn{2}{|c|}{$0\leq\delta<2\pi/3$, $2\pi/3\leq\alpha<4\pi/3$ and $4\pi/3\leq\beta<2\pi$}\\
\hline\hline
$0.7\lesssim\left|m_{ee}/m_{e\mu}\right|\lesssim5\times10^2$ & $0.5\lesssim\left|m_{ee}/m_{\mu\mu}\right|\lesssim4\times10^2$ \\ 
\hline
$1.2\times10^{-3}\lesssim\left|m_{e\tau}/m_{\mu\tau}\right|\lesssim4$ & $1.8\times10^{-6}\lesssim\left|m_{e\tau}^2/(m_{ee}m_{\mu\tau})\right|\lesssim4$ \\
\hline
\multicolumn{2}{c}{~}\\
\hline
\multicolumn{2}{|c|}{$0\leq(\delta,\beta)<2\pi/3$ and $4\pi/3\leq\alpha<2\pi$}\\
\hline\hline
$0.2\lesssim\left|m_{ee}/m_{e\mu}\right|\lesssim8\times10^2$ & $0.3\lesssim\left|m_{ee}/m_{\mu\mu}\right|\lesssim8\times10^2$  \\ 
\hline
$2\times10^{-2}\lesssim\left|m_{e\tau}/m_{\mu\tau}\right|\lesssim4.1$ & $4\times10^{-4}\lesssim\left|m_{e\tau}^2/(m_{ee}m_{\mu\tau})\right|\lesssim6$ \\
\hline
\multicolumn{2}{c}{~}\\
\hline
\multicolumn{2}{|c|}{$0\leq\delta<2\pi/3$, $4\pi/3\leq\alpha<2\pi$ and $2\pi/3\leq\beta<4\pi/3$}\\
\hline\hline
$0.3\lesssim\left|m_{ee}/m_{e\mu}\right|\lesssim8\times10^2$ & $0.3\lesssim\left|m_{ee}/m_{\mu\mu}\right|\lesssim2.2\times10^2$  \\ 
\hline
$4\times10^{-3}\lesssim\left|m_{e\tau}/m_{\mu\tau}\right|\lesssim5$ & $2\times10^{-6}\lesssim\left|m_{e\tau}^2/(m_{ee}m_{\mu\tau})\right|\lesssim2$ \\
\hline
\multicolumn{2}{c}{~}\\
\hline
\multicolumn{2}{|c|}{$0\leq\delta<2\pi/3$ and $4\pi/3\leq(\alpha,\beta)<2\pi$}\\
\hline\hline
$0.3\lesssim\left|m_{ee}/m_{e\mu}\right|\lesssim5\times10^2$ & $0.3\lesssim\left|m_{ee}/m_{\mu\mu}\right|\lesssim2\times10^2$  \\ 
\hline
$9\times10^{-3}\lesssim\left|m_{e\tau}/m_{\mu\tau}\right|\lesssim4$ & $2\times10^{-5}\lesssim\left|m_{e\tau}^2/(m_{ee}m_{\mu\tau})\right|\lesssim6$ \\
\hline
\end{tabular}
\end{center}
\caption{Values of the renormalization invariants for the IH.}
\label{tab4}
\end{table}
\begin{table}
\begin{center}
\begin{tabular}{|c|c|}
\hline
\multicolumn{2}{|c|}{$2\pi/3\leq\delta<4\pi/3$ and $0\leq(\alpha,\beta)<2\pi/3$}\\
\hline\hline
$0.4\lesssim\left|m_{ee}/m_{e\mu}\right|\lesssim1.8\times10^3$ & $0.3\lesssim\left|m_{ee}/m_{\mu\mu}\right|\lesssim2\times10^2$  \\ 
\hline
$7\times10^{-3}\lesssim\left|m_{e\tau}/m_{\mu\tau}\right|\lesssim5.5$ & $5\times10^{-5}\lesssim\left|m_{e\tau}^2/(m_{ee}m_{\mu\tau})\right|\lesssim9$ \\
\hline
\multicolumn{2}{c}{~}\\
\hline
\hline
\multicolumn{2}{|c|}{$2\pi/3\leq(\delta,\beta)<4\pi/3$ and $0\leq\alpha<2\pi/3$}\\
\hline\hline
$0.4\lesssim\left|m_{ee}/m_{e\mu}\right|\lesssim8\times10^2$ & $0.3\lesssim\left|m_{ee}/m_{\mu\mu}\right|\lesssim1.2\times10^2$  \\ 
\hline
$2\times10^{-3}\lesssim\left|m_{e\tau}/m_{\mu\tau}\right|\lesssim6$ & $1.5\times10^{-6}\lesssim\left|m_{e\tau}^2/(m_{ee}m_{\mu\tau})\right|\lesssim5$ \\
\hline
\multicolumn{2}{c}{~}\\
\hline
\multicolumn{2}{|c|}{$2\pi/3\leq\delta<4\pi/3$, $0\leq\alpha<2\pi/3$ and $4\pi/3\leq\beta<2\pi$}\\
\hline\hline
$0.4\lesssim\left|m_{ee}/m_{e\mu}\right|\lesssim5\times10^2$ & $0.3\lesssim\left|m_{ee}/m_{\mu\mu}\right|\lesssim2\times10^2$  \\ 
\hline
$2\times10^{-3}\lesssim\left|m_{e\tau}/m_{\mu\tau}\right|\lesssim5.5$ & $2\times10^{-6}\lesssim\left|m_{e\tau}^2/(m_{ee}m_{\mu\tau})\right|\lesssim5$ \\
\hline
\multicolumn{2}{c}{~}\\
\hline
\multicolumn{2}{|c|}{$2\pi/3\leq(\delta,\alpha)<4\pi/3$ and $0\leq\beta<2\pi/3$}\\
\hline\hline
$0.7\lesssim\left|m_{ee}/m_{e\mu}\right|\lesssim4\times10^2$ & $0.5\lesssim\left|m_{ee}/m_{\mu\mu}\right|\lesssim2\times10^2$  \\ 
\hline
$1.8\times10^{-3}\lesssim\left|m_{e\tau}/m_{\mu\tau}\right|\lesssim6$ & $1.1\times10^{-6}\lesssim\left|m_{e\tau}^2/(m_{ee}m_{\mu\tau})\right|\lesssim5$ \\
\hline
\multicolumn{2}{c}{~}\\
\hline
\multicolumn{2}{|c|}{$2\pi/3\leq(\delta,\alpha,\beta)<4\pi/3$}\\
\hline\hline
$0.7\lesssim\left|m_{ee}/m_{e\mu}\right|\lesssim8\times10^2$ & $0.5\lesssim\left|m_{ee}/m_{\mu\mu}\right|\lesssim1.8\times10^2$  \\ 
\hline
$1.8\times10^{-3}\lesssim\left|m_{e\tau}/m_{\mu\tau}\right|\lesssim6.5$ & $1.3\times10^{-6}\lesssim\left|m_{e\tau}^2/(m_{ee}m_{\mu\tau})\right|\lesssim5$  \\
\hline
\multicolumn{2}{c}{~}\\
\hline
\multicolumn{2}{|c|}{$2\pi/3\leq(\delta,\alpha)<4\pi/3$ and $4\pi/3\leq\beta<2\pi$}\\
\hline\hline
$0.7\lesssim\left|m_{ee}/m_{e\mu}\right|\lesssim3\times10^3$ & $0.5\lesssim\left|m_{ee}/m_{\mu\mu}\right|\lesssim5\times10^2$ \\ 
\hline
$3.2\times10^{-3}\lesssim\left|m_{e\tau}/m_{\mu\tau}\right|\lesssim5.2$ & $10^{-5}\lesssim\left|m_{e\tau}^2/(m_{ee}m_{\mu\tau})\right|\lesssim5$ \\
\hline
\multicolumn{2}{c}{~}\\
\hline
\multicolumn{2}{|c|}{$2\pi/3\leq\delta<4\pi/3$, $4\pi/3\leq\alpha<2\pi$ and $0\leq\beta<2\pi/3$}\\
\hline\hline
$0.4\lesssim\left|m_{ee}/m_{e\mu}\right|\lesssim3\times10^2$ & $0.3\lesssim\left|m_{ee}/m_{\mu\mu}\right|\lesssim2\times10^2$  \\ 
\hline
$3\times10^{-3}\lesssim\left|m_{e\tau}/m_{\mu\tau}\right|\lesssim4$ & $10^{-5}\lesssim\left|m_{e\tau}^2/(m_{ee}m_{\mu\tau})\right|\lesssim5$ \\
\hline
\multicolumn{2}{c}{~}\\
\hline
\multicolumn{2}{|c|}{$2\pi/3\leq(\delta,\beta)<4\pi/3$ and $4\pi/3\leq\alpha<2\pi$}\\
\hline\hline
$0.4\lesssim\left|m_{ee}/m_{e\mu}\right|\lesssim8\times10^2$ & $0.3\lesssim\left|m_{ee}/m_{\mu\mu}\right|\lesssim3\times10^2$  \\ 
\hline
$2\times10^{-3}\lesssim\left|m_{e\tau}/m_{\mu\tau}\right|\lesssim5.5$ & $10^{-6}\lesssim\left|m_{e\tau}^2/(m_{ee}m_{\mu\tau})\right|\lesssim5$ \\
\hline
\multicolumn{2}{c}{~}\\
\hline
\multicolumn{2}{|c|}{$2\pi/3\leq\delta<4\pi/3$ and $4\pi/3\leq(\alpha,\beta)<2\pi$}\\
\hline\hline
$0.4\lesssim\left|m_{ee}/m_{e\mu}\right|\lesssim8\times10^2$ & $0.2\lesssim\left|m_{ee}/m_{\mu\mu}\right|\lesssim5\times10^2$  \\ 
\hline
$2\times10^{-3}\lesssim\left|m_{e\tau}/m_{\mu\tau}\right|\lesssim5.5$ & $2\times10^{-6}\lesssim\left|m_{e\tau}^2/(m_{ee}m_{\mu\tau})\right|\lesssim5$ \\
\hline
\end{tabular}
\end{center}
\caption{Values of the renormalization invariants for the IH.}
\label{tab5}
\end{table}
\begin{table}
\begin{center}
\begin{tabular}{|c|c|}
\hline
\multicolumn{2}{|c|}{$4\pi/3\leq\delta<2\pi$ and $0\leq(\alpha,\beta)<2\pi/3$}\\
\hline\hline
$0.3\lesssim\left|m_{ee}/m_{e\mu}\right|\lesssim5\times10^2$ & $0.3\lesssim\left|m_{ee}/m_{\mu\mu}\right|\lesssim4\times10^2$  \\ 
\hline
$10^{-3}\lesssim\left|m_{e\tau}/m_{\mu\tau}\right|\lesssim4$ & $1.8\times10^{-5}\lesssim\left|m_{e\tau}^2/(m_{ee}m_{\mu\tau})\right|\lesssim5$ \\
\hline
\multicolumn{2}{c}{~}\\
\hline
\hline
\multicolumn{2}{|c|}{$4\pi/3\leq\delta<2\pi$, $0\leq\alpha<2\pi/3$ and $2\pi/3\leq\beta<4\pi/3$}\\
\hline\hline
$0.3\lesssim\left|m_{ee}/m_{e\mu}\right|\lesssim9\times10^2$ & $0.3\lesssim\left|m_{ee}/m_{\mu\mu}\right|\lesssim70$  \\ 
\hline
$4\times10^{-3}\lesssim\left|m_{e\tau}/m_{\mu\tau}\right|\lesssim4$ & $8\times10^{-6}\lesssim\left|m_{e\tau}^2/(m_{ee}m_{\mu\tau})\right|\lesssim5$ \\
\hline
\multicolumn{2}{c}{~}\\
\hline
\multicolumn{2}{|c|}{$4\pi/3\leq(\delta,\beta)<2\pi$ and $0\leq\alpha<2\pi/3$}\\
\hline\hline
$0.3\lesssim\left|m_{ee}/m_{e\mu}\right|\lesssim3.5\times10^2$ & $0.3\lesssim\left|m_{ee}/m_{\mu\mu}\right|\lesssim6\times10^2$  \\ 
\hline
$2.2\times10^{-2}\lesssim\left|m_{e\tau}/m_{\mu\tau}\right|\lesssim4$ & $2\times10^{-4}\lesssim\left|m_{e\tau}^2/(m_{ee}m_{\mu\tau})\right|\lesssim5$ \\
\hline
\multicolumn{2}{c}{~}\\
\hline
\multicolumn{2}{|c|}{$4\pi/3\leq\delta<2\pi$, $2\pi/3\leq\alpha<4\pi/3$ and $0\leq\beta<2\pi/3$}\\
\hline\hline
$0.6\lesssim\left|m_{ee}/m_{e\mu}\right|\lesssim6\times10^2$ & $0.5\lesssim\left|m_{ee}/m_{\mu\mu}\right|\lesssim5\times10^2$  \\ 
\hline
$1.2\times10^{-3}\lesssim\left|m_{e\tau}/m_{\mu\tau}\right|\lesssim4$ & $1.2\times10^{-6}\lesssim\left|m_{e\tau}^2/(m_{ee}m_{\mu\tau})\right|\lesssim5$ \\
\hline
\multicolumn{2}{c}{~}\\
\hline
\multicolumn{2}{|c|}{$4\pi/3\leq\delta<2\pi$ and $2\pi/3\leq(\alpha,\beta)<4\pi/3$}\\
\hline\hline
$0.6\lesssim\left|m_{ee}/m_{e\mu}\right|\lesssim7\times10^2$ & $0.5\lesssim\left|m_{ee}/m_{\mu\mu}\right|\lesssim10^2$  \\ 
\hline
$3\times10^{-3}\lesssim\left|m_{e\tau}/m_{\mu\tau}\right|\lesssim6$ & $1.2\times10^{-6}\lesssim\left|m_{e\tau}^2/(m_{ee}m_{\mu\tau})\right|\lesssim5$  \\
\hline
\multicolumn{2}{c}{~}\\
\hline
\multicolumn{2}{|c|}{$4\pi/3\leq(\delta,\beta)<2\pi$ and $2\pi/3\leq\alpha<4\pi/3$}\\
\hline\hline
$0.6\lesssim\left|m_{ee}/m_{e\mu}\right|\lesssim8\times10^2$ & $0.4\lesssim\left|m_{ee}/m_{\mu\mu}\right|\lesssim2\times10^2$ \\ 
\hline
$1.2\times10^{-3}\lesssim\left|m_{e\tau}/m_{\mu\tau}\right|\lesssim7.5$ & $10^{-6}\lesssim\left|m_{e\tau}^2/(m_{ee}m_{\mu\tau})\right|\lesssim5$ \\
\hline
\multicolumn{2}{c}{~}\\
\hline
\multicolumn{2}{|c|}{$4\pi/3\leq(\delta,\alpha)<2\pi$ and $0\leq\beta<2\pi/3$}\\
\hline\hline
$0.2\lesssim\left|m_{ee}/m_{e\mu}\right|\lesssim3.5\times10^2$ & $0.2\lesssim\left|m_{ee}/m_{\mu\mu}\right|\lesssim3\times10^2$  \\ 
\hline
$3\times10^{-3}\lesssim\left|m_{e\tau}/m_{\mu\tau}\right|\lesssim4$ & $10^{-5}\lesssim\left|m_{e\tau}^2/(m_{ee}m_{\mu\tau})\right|\lesssim5$ \\
\hline
\multicolumn{2}{c}{~}\\
\hline
\multicolumn{2}{|c|}{$4\pi/3\leq(\delta,\alpha)<2\pi$ and $2\pi/3\leq\beta<4\pi/3$}\\
\hline\hline
$0.3\lesssim\left|m_{ee}/m_{e\mu}\right|\lesssim10^3$ & $0.2\lesssim\left|m_{ee}/m_{\mu\mu}\right|\lesssim5\times10^2$  \\ 
\hline
$3\times10^{-3}\lesssim\left|m_{e\tau}/m_{\mu\tau}\right|\lesssim6$ & $3\times10^{-6}\lesssim\left|m_{e\tau}^2/(m_{ee}m_{\mu\tau})\right|\lesssim5$ \\
\hline
\end{tabular}
\end{center}
\caption{Values of the renormalization invariants for the IH.}
\label{tab6}
\end{table}


\end{document}